\magnification=\magstep1
\def\bs{\bigskip}
\def\ni{\noindent}

\centerline{\bf Comment on "Brown Dwarfs, Quark Stars, and Quark-Hadron}

\centerline{\bf Phase Transition"}

\bs
\bs
\bs

\centerline{S. Kubis and M. Kutschera}

\centerline{H. Niewodnicza\'nski Institute of Nuclear Physics}

\centerline {ul. Radzikowskiego 152, 31-342 Krak\'ow, Poland}

\bs\bs\bs

In a recent Letter, Cottingham, Kalafatis and Vinh Mau [1]
studied, within the Lee-Wick model, the formation of quark 
stars in the cosmological quark-hadron phase transition. They
conclude that the formation of quark stars with the solar baryon
number $N_{\odot}\sim 10^{57}$ requires a high degree of
supercooling which can be achieved with a reasonable choice of
the parameters of the Lee-Wick model. 

We show below that the cosmological quark-hadron phase
transition in the Lee-Wick model with such a
high degree of supercooling  cannot be
completed and thus no quark stars can be formed in this
scenario. Conclusions of Ref.[1] are based on the
assumption that the expansion of the Universe is dominated by
radiation during the phase transition, with the scale factor
$R(t)\sim \sqrt{t}$. This assumption is, however, unjustified. 

For cosmological application, the potential energy of the
Lee-Wick model has to be chosen properly. The very small value of
the cosmological constant requires that the energy of the true
vacuum, $\sigma=\sigma_{vac}$, is essentially zero,
$U(\sigma_{vac})=0$. Then the energy of the false vacuum,
$\sigma=0$, is $U(0)=B>0$. With this choice of $U$, when the
degree of supercooling is as high as required in Ref.[1], the
phase transition is slow and the expansion of the Universe
becomes soon dominated by the vacuum energy $B$. As a result,
bubbles of a new phase do not percolate.

To show that the phase transition is slow we
compare the bubble nucleation rate, $\Gamma(t)$, and the expansion
rate, $H(t)$, by calculating the dimensionless
quantity $\epsilon(t)=\Gamma(t)/H(t)^4$ [2]. The expansion rate,
$H(t)={\dot R(t)}/R(t)$, satisfies the equation

$$  H^2={8\pi \over 3}G({\rho_c \over R^4}+B),$$

\ni where $\rho_c$ is the energy density of massless particles
at the critical temperature $T_c$, $\rho_c=g_{eff}aT_c^4/2$,
with $g_{eff}\approx 32$ in the Lee-Wick model. 
Normalizing the scale factor at $t_c$, $R(t_c)=1$, where time
$t_c$ corresponds to the critical temperature $T_c$, we find the
solution 

$$ R(t)={1 \over{\sqrt{2\lambda}}}e^{-\alpha t/t_c}\sqrt{e^{4\alpha
t/t_c}-1},$$ 

\ni where $\lambda^2=B/\rho_c\approx 1/5$ and
$\alpha=ln({\sqrt{\lambda^2+1}+\lambda})/2\approx 0.2$. 

 The bubble nucleation rate, $\Gamma(t)$, for adiabatic
expansion, $TR=T_c$,  is calculated using Eq.(6) and Eq.(7) of
Ref.[1] with $T(t)/T_c=1/R(t)$. As shown in [1], 
$\Gamma(t)$ is always strongly suppressed except in a narrow
time interval around time $t_n$ when almost all the bubbles are
nucleated. Correspondingly,
the values of $\epsilon(t)$ are always lower than
the maximum value $\epsilon_{max}(t_n)=7.5 \times 10^{-10}$
occurring at $t_n\approx 2t_c$ when the 
scale factor is $R(t_n)\approx \sqrt{3}$.
The values of $\epsilon(t)$ are always below the percolation
threshold [2]. The nucleation of bubbles through the quantum
tunneling is negligible here.   

The density of nucleated bubbles at $t_n$ is of the order of
$10^{-30}cm^{-3}$ and a typical distance 
between bubbles is of the order of $10^{10} cm$. It
exceeds the Hubble distance $1/H(t_n)$ by
a factor of the order of $1000$. 
The separation of neighbouring bubbles is so big that there is no chance
for them to coalesce during the intermediate evolution, when
$R(t)$ approaches the exponential behaviour. After the exponential 
expansion sets in, the bubble density vanishes exponentially.
Similarly as in the old inflationary model due to Guth,
the phase transition is never completed [2]. 

The Lee-Wick model with a high degree of supercooling leads to the
unphysical scenario for the cosmological quark-hadron phase transition.
However, moderate changes of the Lee-Wick model
parameters can drastically lower the degree of supercooling. The
phase transition with the low degree of supercooling is
quickly completed and no unphysical behaviour, found for
a strongly supercooled system, is encountered. In this case, however,
only smaller aggregates of quark matter, with the baryon number
$N_B<<10^{-5}N_{\odot}$, could be formed.

This research is partially supported by the Polish State Committee
for Scientific Research (KBN), grants  2 P03B 083 08 and 2 P03D
001 09.

\bs
\bs
\ni REFERENCES

\ni [1]~W.N.Cottingham, D. Kalafatis, and R. Vinh Mau, Phys.
Rev. Lett. {\bf 73}, 1328 (1994). 

\ni [2]~A.H. Guth, and E.J. Weinberg, Nucl. Phys. {\bf B 212},
321 (1983).
\end